\documentclass[prapplied,reprint,superscriptaddress]{revtex4-2}
\usepackage{graphicx}
\usepackage{bm}
\usepackage[linktocpage=true,colorlinks=true,urlcolor=blue,linkcolor=red,citecolor = blue]{hyperref}
\usepackage{amsfonts}
\usepackage{amsmath}
\usepackage{amssymb}
\usepackage{braket}
\usepackage{newfloat}
\usepackage{siunitx}
\usepackage{physics}
\usepackage{mathtools}
\usepackage{xcolor}
\usepackage{comment}
\usepackage{array}
\newcommand{\red}[1]{\textcolor{red}{#1}}

\usepackage[T1]{fontenc}
\usepackage{tabularx}
\usepackage{booktabs}
\usepackage{makecell}
\usepackage{booktabs}
\usepackage{placeins}
\newcolumntype{W}[1]{>{\centering\arraybackslash}p{#1}}

\newcolumntype{Y}{>{\centering\arraybackslash}m{3.2cm}} 
\newcolumntype{Z}{>{\centering\arraybackslash}m{1.8cm}}   

\graphicspath{ {Figures/} }

\begin{document}

\title{Enhancing Valley Splitting in Si/SiGe Wiggle Well Quantum Dots through Integrated Stressor–Gate Architectures}

\author{E. S. Joseph}
\affiliation{Department of Physics, University of Wisconsin-Madison, Madison, Wisconsin 53706, USA}
\affiliation{Q-NEXT, Argonne National Laboratory, Lemont, Illinois 60439, USA}

\author{Talise Oh}
\affiliation{Department of Physics, University of Wisconsin-Madison, Madison, Wisconsin 53706, USA}

\author{D. E. Savage}
\affiliation{Department of Materials Science and Engineering, University of Wisconsin-Madison, Madison, Wisconsin 53706, USA}

\author{D. S. Stone}
\affiliation{Department of Materials Science and Engineering, University of Wisconsin-Madison, Madison, Wisconsin 53706, USA}

\author{DeAnna Campbell}
\affiliation{Sandia National Laboratories, Albuquerque, NM 87185, USA}

\author{Benjamin D. Woods}
\affiliation{Department of Physics, University of Wisconsin-Madison, Madison, Wisconsin 53706, USA}

\author{Mark Friesen}
\affiliation{Department of Physics, University of Wisconsin-Madison, Madison, Wisconsin 53706, USA}

\author{M. A. Eriksson}
\affiliation{Department of Physics, University of Wisconsin-Madison, Madison, Wisconsin 53706, USA}

\begin{abstract}
Engineering large and reproducible valley splittings is essential for the scalable operation of silicon spin qubits. In Si/SiGe quantum dot devices, valley splitting arises predominantly from atomic-scale disorder and is therefore inherently random, motivating approaches that introduce a large deterministic contribution. Strain-assisted Wiggle Wells provide a route to deterministic valley splitting by combining an oscillatory Ge concentration in the Si well with shear strain; however, generating the required strain in a manner compatible with large-scale architectures remains a challenge. Here, we propose patterned stressed thin films as a scalable approach for generating shear strain in the quantum well of Si/SiGe quantum dot devices. Finite-element simulations show that compressively stressed TiN films can generate shear strains exceeding 0.1\% at the quantum well, sufficient for deterministic valley splittings of hundreds of $\mu \text{eV}$ in strain-assisted Wiggle Wells. We show that this strain can extend across multiple quantum dots and is maintained in realistic gate geometries, providing a fabrication-compatible approach for scaling deterministic valley splitting to quantum dot arrays.
\end{abstract}

\maketitle

\section{Introduction}

\begin{figure}[t]
\begin{center}
\includegraphics[width=0.47\textwidth]{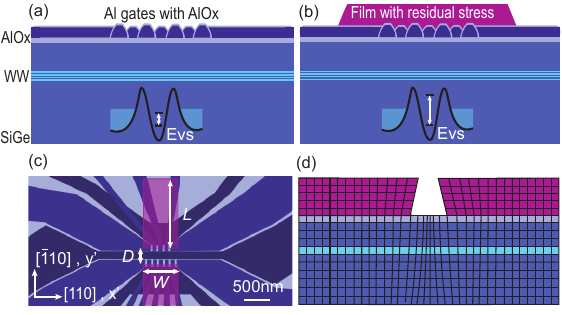}
\end{center}
\vspace{-0.5cm}
\hspace{-1cm}
\caption{Schematic illustration of how strain can be applied to a quantum dot formed in a Wiggle Well heterostructure.
(a) Cross-sectional schematic of a Si/SiGe heterostructure with a Wiggle Well and no applied stressor, showing a resulting valley splitting $E_{\mathrm{VS}}$ (inset). 
(b) Cross-sectional schematic of the same heterostructure with a stressor fabricated on top of the device, intended to enhance the shear strain $\varepsilon_{xy}$ and increase the valley splitting $E_{\mathrm{VS}}$. 
(c) Top-down view of the device layout with overlapping gate architecture and an overlying stressor, where the dark blue and lighter blue colors denote the first and second gate layers, respectively; the third gate layer is omitted for clarity. The stressors are centered near the dot region of the device and are intentionally oriented along the $[\bar{1}10]$ lattice direction to induce shear strain. The stressor length ($L$), width ($W$), and spacing ($D$) are indicated. 
(d) Cross-sectional schematic illustrating the strain distribution generated by surface stressors in the Si/SiGe heterostructure. 
}
\vspace{-1mm}
\label{FIG_Intro}
\end{figure}

Si/SiGe quantum dots are a leading platform for quantum computing due to their compatibility with existing semiconductor technology and long spin coherence times~\cite{Loss1998QuantumDots,Hanson2007SpinsDots,Burkard2023SemiconductorQubits}. However, their performance is limited by the small energy separation between ground and excited valley states, known as the valley splitting $E_\text{VS}$~\cite{Friesen2007_2, Zwanenburg2013SiliconElectronics}. Random alloy disorder in the quantum well causes $E_\text{VS}$ to vary significantly from dot to dot, typically in a range from tens to several hundred $\mu$eV~\cite{Losert2024StrategiesDisorder, PaqueletWuetz2023ReducingWells, Lima2024ValleyDot, Neyens2018TheWells}. Incorporating Ge into the well increases alloy disorder and can enhance the average $E_\text{VS}$, but pockets of low valley splitting still persist~\cite{Marcks2025ValleyGermanium}. A central challenge for scalable Si/SiGe quantum computing is therefore achieving a deterministic and spatially uniform enhancement of $E_\text{VS}$ across devices.

The Wiggle Well (WW) approach addresses this challenge by introducing Ge concentration oscillations -- “wiggles” -- into the Si quantum well, characterized by a wavelength $\lambda$. A short wavelength of $\lambda = 0.32$~nm has been theoretically shown to couple valleys within a single Brillouin zone, yielding deterministic valley splittings as large as $E_\text{VS} \approx 1~$meV~\cite{McJunkin:2022p7777, Feng2022EnhancedConcentration}. However, this wavelength corresponds to approximately two atomic monolayers, making such structures extremely difficult to grow. A longer wavelength Wiggle Well, with $\lambda = 1.8$~nm, could couple valleys in neighboring Brillouin zones, but this coupling is suppressed by the nonsymmorphic screw symmetry of the Si crystal. To overcome this symmetry constraint, the strain-assisted Wiggle Well (STRAWW) approach was proposed, combining long-wavelength Ge oscillations with shear strain to enable inter-zone valley coupling \cite{Woods2024CouplingOscillations}. 

Importantly, the long-wavelength Wiggle Well component of this proposal has already been experimentally demonstrated using both chemical vapor deposition, and more recently, molecular beam epitaxy, showing that these structures are compatible with a variety of epitaxial growth techniques~\cite{McJunkin:2022p7777, Gradwohl2025EnhancedSegregation}. In the CVD-grown structure of Ref.~\cite{McJunkin:2022p7777}, the Ge concentration oscillated between approximately 0\% and 9\% with a wavelength of $\lambda=1.7$~nm, very near the long-wavelength (inter-Brillouin-zone) coupling condition. These quantum wells exhibited large and tunable valley splittings ranging from 53-239~$\mu \text{eV}$~\cite{McJunkin:2022p7777}. Notably however, these measurements were performed without any intentionally introduced shear strain, suggesting that disorder-related mechanisms provided the dominant contribution to the observed valley splitting. When combined with a substantial but experimentally realistic shear strain of approximately 0.1\%, this Wiggle Well structure is predicted to exhibit an average deterministic valley splitting enhancement of $E_\text{VS} \geq 100~\mu \text{eV}$~\cite{Woods2024CouplingOscillations}. While it is known that electrodes fabricated on the surface of the heterostructure produce strain fields in the material, many gate structures generate shear strains on the order of 0.01\% \cite{Thorbeck2015FormationNanostructures, Park2016Electrode-stress-inducedDevices, Corley-Wiciak2023LatticeDevices, Frink2025ReducingStacking}, or sometimes a bit more ($\sim$0.025\% \cite{Corley-Wiciak2022NanoscaleDevice}), which is a factor 4-10~times smaller than the strain targeted for STRAWW. 
On the other hand, some micromagnets have been found to impart sufficient shear strain magnitudes of $>$0.4\% \cite{Frink2025ReducingStacking}.
Thus, while the long-wavelength Wiggle Well structure has already been experimentally demonstrated, generating sufficiently large shear strain requires careful consideration and currently remains an open challenge for realizing the STRAWW proposal.

In this work, we investigate the use of surface-deposited stressors to generate the shear strain required for Wiggle Well quantum dot devices. Through finite-element simulations, we determine how the induced strain depends on the stressor geometry, intrinsic stress, and placement relative to the quantum well, and we identify the fabrication tolerances required to achieve the desired strain at the quantum dot location. We further examine the integration of this approach with realistic gate architectures and show that the resulting strain fields remain compatible with realistic device fabrication. For representative stressor geometries and experimentally achievable material parameters, the induced shear strain enhances the deterministic component of the valley splitting to values on the order of hundreds of $\mu\mathrm{eV}$.

Figure~\ref{FIG_Intro} provides an overview of the strain-engineering approach investigated in this work. Without an applied stressor, the valley splitting relies on contributions from the quantum-well interfaces, alloy disorder, and the long-wavelength Wiggle Well Ge oscillations [Fig.~\ref{FIG_Intro}(a)]. A surface stressor generates shear strain in the quantum well, adding a deterministic strain-assisted contribution to the valley splitting [Fig.~\ref{FIG_Intro}(b)]. The stressor geometry and placement relative to the quantum-dot gate architecture are shown in Fig.~\ref{FIG_Intro}(c), while Fig.~\ref{FIG_Intro}(d) shows a schematic strain field, which we flesh out in the simulations presented in the remainder of the paper.

To investigate the shear strain imparted to the quantum dot region, we simulate a series of device geometries. For clarity, we begin with a simplified Si/SiGe heterostructure containing a quantum well and a single surface stressor, neglecting the complications introduced by the underlying gate metal layers. We then investigate how pairs of stressors can be used to enhance the shear strain at the quantum dot location. Next, we examine stressors integrated above a simplified gate geometry to isolate the effect of the gate stack on strain transfer. Finally, we incorporate these design principles into a realistic quadruple-dot device shown, to evaluate the feasibility of implementing this approach in an experimentally relevant architecture.

\section{Results}

\subsection{Single stressor}
\label{Sec:Single_Stressor}

\begin{figure}[t!]
\centering
\includegraphics[width=0.92\columnwidth]{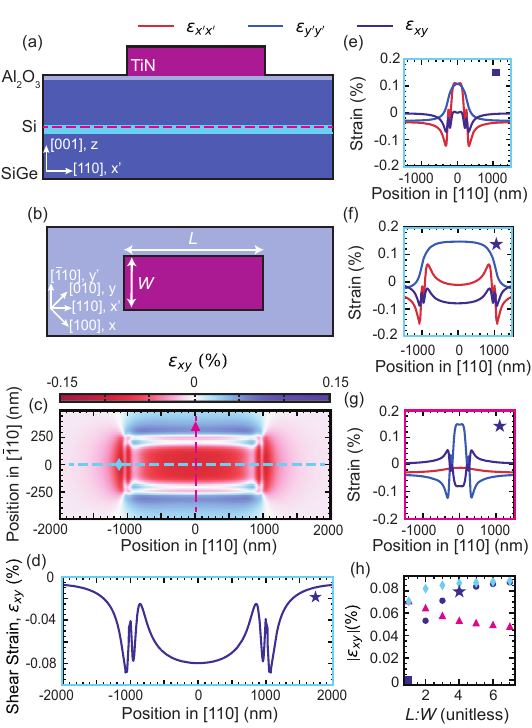}
\caption{Shear strain produced by a single rectangular stressor. 
(a) Cross-sectional schematic (not shown to scale) of a Si/SiGe heterostructure with a single surface stressor and a conventional Si quantum well. 
Throughout this work, we consider a \SI{2}{\micro\meter} layer of strain-relaxed Si$_{0.7}$Ge$_{0.3}$, a \SI{9}{\nano\meter} Si quantum well, a \SI{30}{\nano\meter} Si$_{0.7}$Ge$_{0.3}$ spacer layer, a \SI{10}{\nano\meter} insulating layer of Al$_2$O$_3$, and TiN stressor(s) of thickness $t_{\mathrm{stressor}}=~$\SI{30}{\nano\meter}, unless otherwise indicated.
The dashed line indicates the location in the quantum well where the strain is evaluated. 
(b) Top-down schematic of the stressor, with its long axis parallel to the $[110]$ crystallographic direction (defined as $\hat{x}'$). 
The length ($L$) and width ($W$) of the stressor are indicated. 
(c) A 2D map of the shear strain $\varepsilon_{xy}$, as defined in Eq.~(\ref{eq:shear_transform}).
Two linecuts will be considered in the following panels, defined by the dashed light blue and fuchsia dashed lines.
(This color coding is also used below.)
(d) A linecut of $\varepsilon_{xy}$ along the [110] direction, corresponding to the blue dashed line in (c). 
(e,f) Linecuts along the $[110]$ direction, corresponding to the blue dashed line in (c), for (e) the 1:1 aspect ratio, and (f) the 4:1 aspect ratio.
Here, the different curves correspond to $\varepsilon_{x'x'}$ (red), $\varepsilon_{y'y'}$ (lighter blue), and $\varepsilon_{xy}$ (darker blue).
Panel (d) shows a blown-up view of the latter plot.
(g) Linecuts along the $[\bar{1}10]$ or $\hat{y}'$ direction, corresponding to the pink dashed line in (c). 
(h) $\varepsilon_{xy}$ evaluated at three locations in the device, corresponding to the center of the stressor (dark blue circle), the center of the short edge of the stressor (light blue diamond), and the center of the long edge of the stressor (pink triangle).
Results are shown as functions of the stressor aspect ratio $L$:$W$. 
The dark blue square denotes the 1:1 aspect ratio and the dark blue star denotes the 4:1 aspect ratio.
}
\label{FIG_SingleStressor}
\end{figure}

We first consider a simple geometry consisting of a Si/SiGe heterostructure with a single stressor patterned on the surface, as depicted in Figs.~\ref{FIG_SingleStressor}(a) and \ref{FIG_SingleStressor}(b). 
A cross-sectional schematic is shown in Fig.~\ref{FIG_SingleStressor}(a). From bottom to top, the simulated heterostructure consists of a \SI{2}{\micro\meter} layer of strain-relaxed Si$_{0.7}$Ge$_{0.3}$, a \SI{9}{\nano\meter} Si quantum well, a \SI{30}{\nano\meter} Si$_{0.7}$Ge$_{0.3}$ spacer layer, and a \SI{10}{\nano\meter} insulating layer of Al$_2$O$_3$. We assume that all interfaces are epitaxially grown, and that the Si quantum well remains fully strained: that is, it does not relax during growth. As a result, the Si layer experiences biaxial tensile strain due to the larger lattice constant of the surrounding SiGe. 
This strain lifts the degeneracy between the in-plane $x$ and $y$ valleys and the out-of plane $z$ valleys, with the $z$ valleys having lower energy. However, because biaxial tensile strain preserves the in-plane symmetry of the quantum well, it does not lift the remaining degeneracy between the two $z$ valleys. 
Unless otherwise specified, the heterostructure of Fig.~\ref{FIG_SingleStressor} is assumed throughout this work. 
The strain and the resulting energy variations are evaluated in the plane of the two-dimensional electron gas (2DEG), which is assumed to reside \SI{1.5}{\nano\meter} below the top interface of the quantum well.

We now consider the effect of a single deposited rectangular stressor with length $L$ and width $W$, oriented along the $[110]$ and $[\bar{1}10]$ crystallographic directions, as illustrated in Fig.~\ref{FIG_SingleStressor}(b).
We define the axes for this system as $\hat{x}'$ and $\hat{y}'$, as shown in the insets of the figure, while the crystallographic axes are defined as $\hat{x}$ and $\hat{y}$.
For convenience, throughout this work we perform our simulations in the $(x',y',z')$ system, where $\hat{z}'=\hat{z}$.
The stressor in Fig.~\ref{FIG_SingleStressor} is taken to be a \SI{30}{\nano\meter} TiN film with a compressive residual stress of $-2.6~\text{GPa}$ ~\cite{Corley-Wiciak2023LatticeDevices}.
The simulations are performed using COMSOL Multiphysics software \cite{COMSOLABV.5.6}, and they also include the effects of thermal contractions, as detailed in Appendix~\ref{appendix:COMSOL}.

Results of the single-stressor strain simulations are shown in Figs.~\ref{FIG_SingleStressor}(c)-\ref{FIG_SingleStressor}(h).
In Figs.~\ref{FIG_SingleStressor}(e)-\ref{FIG_SingleStressor}(g), the results are plotted along the linecuts shown as dashed lines in Fig.~\ref{FIG_SingleStressor}(c), with appropriate color coding of the axes to show the different cuts.
We observe sharp changes in $\varepsilon_{x'x'}$ or $\varepsilon_{y'y'}$ at the edges of the stressor.
Many of the features at the edges of the stressor are manifested as double-bumps, as explained in Appendix~\ref{Sec:EdgeEffect}.

The shear strain $\varepsilon_{xy}$ is of particular interest for valley splitting calculations, and forms the main focus of this work. 
It is important to note that, while shear strain can be computed for any coordinate axes, only $\varepsilon_{xy}$ with respect to the $(x,y,z)$ system has an effect on the valley splitting \cite{Woods2024CouplingOscillations}. 
To compute $\varepsilon_{xy}$, we therefore perform a coordinate transformation as 
\begin{equation}
\varepsilon_{xy}=\frac{\varepsilon_{x'x'} - \varepsilon_{y'y'}}{2}.
\label{eq:shear_transform}
\end{equation}
This formula highlights the importance of orienting devices along the $[110]$ and $[\bar{1}10]$ axes.

The geometrical features of the stressor, especially the aspect ratio $L$:$W$, are also key factors in determining $\varepsilon_{xy}$.
Our simulations show that enhancing $\varepsilon_{xy}$ requires an asymmetry of the stressors, along the $(x',y')$ coordinate axes, as exemplified in Fig.~\ref{FIG_SingleStressor}(b). 
Results for a symmetric 1:1 stressor geometry ($L=W$), lacking this asymmetry, are shown in Fig.~\ref{FIG_SingleStressor}(e), which reports $\varepsilon_{x'x'}$ (red curve) and $\varepsilon_{y'y'}$ (lighter blue curve). 
The resulting shear strain, computed according to Eq.~(\ref{eq:shear_transform}), reveals that $\varepsilon_{xy}=0$ under the center of the stressor (darker blue curve), as expected from symmetry arguments. 
On the other hand, Figs.~\ref{FIG_SingleStressor}(f) and \ref{FIG_SingleStressor}(g) show results for geometries with an asymmetric aspect ratio of 4:1.
With this ratio we often observe larger $\varepsilon_{xy}$ values, although the results are slightly more subtle, as we discuss below.
Because of its importance, a blown-up view of the shear-strain result in Fig.~\ref{FIG_SingleStressor}(f) is shown in Fig.~\ref{FIG_SingleStressor}(d).

Figure~\ref{FIG_SingleStressor}(h) illustrates the behavior of $\varepsilon_{xy}$ for a range of aspect ratios. 
At the center of the stressor (dark blue circles), we find that $\varepsilon_{xy}$ increases with increasing aspect ratios, saturating near the ratio of 4:1.
Similarly, at the center of the shorter edge of the stressor (light blue diamonds) $\varepsilon_{xy}$ also increases with increasing aspect ratios.
In contrast, at the center of the longer edge of the stressor (fuchsia triangles) $\varepsilon_{xy}$ decreases with increasing aspect ratios.
[The diamond and triangle locations are illustrated along the linecuts shown in Fig.~\ref{FIG_SingleStressor}(c).]
Since we focus on obtaining large $\varepsilon_{xy}$ values here, for the remainder of this work, we therefore adopt the saturated 4:1 geometry as the default configuration for different parameter sweeps, unless otherwise specified. 
The results shown in Fig.~\ref{FIG_SingleStressor}(h) highlight our first important conclusion, that the magnitude and the sign of the shear strain near the edges of the sensor depend on whether it is a long edge or a short edge.
The enhanced shear strain near the short edges can be understood as a consequence of the finite geometry of the stressor. Relaxation of the residual film stress occurs near the stressor boundaries, and the resulting strain fields tend to be largest beneath the stressor, or near the edges, and to decrease with distance while moving away from the stressor edge. As the stressor becomes more elongated, the short-edge response increases, while the strain at the center of the long edge decreases, as $L$ grows.
The saturation of the shear strain reflects the asymptotic response corresponding to a long rectangular stressor.

\subsection{Double stressor}
\label{Sec:Double_Stressor}

\begin{figure}[t]
\begin{center}
\includegraphics[width=0.47\textwidth]{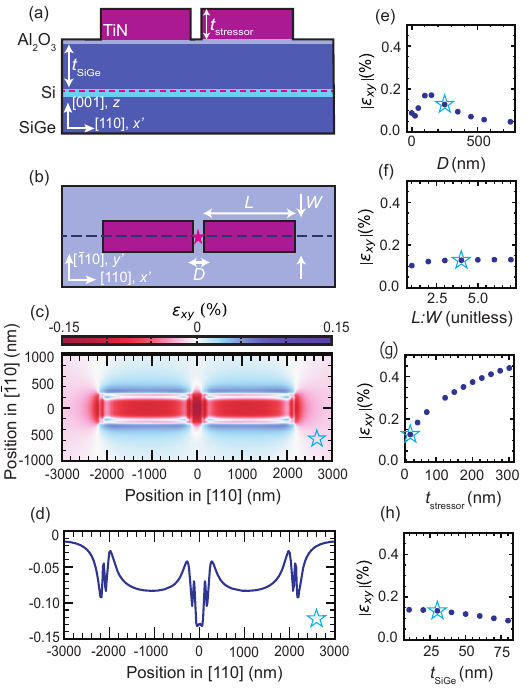}
\end{center}
\vspace{-0.5cm}
\hspace{-1cm}
\caption{Shear-strain simulations of a double-stressor geometry. 
(a) Cross-sectional and (b) top-down schematic views (not shown to scale) of a Si/SiGe heterostructure with two surface stressors. 
Coordinate axes are indicated, along with the stressor thickness ($t_{\mathrm{stressor}}$), SiGe spacer thickness ($t_{\mathrm{SiGe}}$), stressor spacing ($D$), and stressor length ($L$) and width ($W$). The pink star denotes the evaluation location used in (e)-(h). 
(c) A 2D map of the shear strain $\varepsilon_{xy}$.
Here, we assume the device parameters $L=2000$~nm, $W=500$~nm, $D=250$~nm, $t_\text{stressor}=30$~nm, and $t_\text{SiGe}=30$~nm.
(d) A linecut through the data in (c), passing through the center of the stressors, along the dashed line shown in (b).
(e) $\varepsilon_{xy}$ as a function of $D$, with all other parameters held constant. 
(f) $\varepsilon_{xy}$ as a function of $L{:}W$, with all other parameters held constant.
(g) $\varepsilon_{xy}$ as a function of $t_{\mathrm{stressor}}$, with all other parameters held constant.
(h) $\varepsilon_{xy}$ as a function of $t_{\mathrm{SiGe}}$, with all other parameters held constant.  
In (e)-(h), the markers indicated by blue stars all correspond to the same device parameters, which are also the parameters assumed in (c) and (d).
}
\label{FIG_DoubleStressor}
\vspace{-1mm}
\end{figure}

Our goal for designing stressors is to create strain patterns that are large and uniform across a region that could contain dots.
Based on observations in the previous subsection, we therefore consider a configuration consisting of two stressors separated by a distance $D$, as illustrated in Figs.~\ref{FIG_DoubleStressor}(a) and \ref{FIG_DoubleStressor}(b), where we imagine quantum dots located below the pink star, which is centered between the two stressors. This geometry allows us to exploit the edge effect observed for a single stressor and combine the strain contributions constructively. To optimize this effect, we systematically vary geometrical parameters including the stressor spacing $D$, the stressor thickness $t_\text{stressor}$, the aspect ratio $L$:$W$, and the thickness of the SiGe cap $t_\text{SiGe}$. 

Results of our double-stressor simulations are shown in Figs.~\ref{FIG_DoubleStressor}(c)-\ref{FIG_DoubleStressor}(h), where we find that the shear strain at the location of the pink star in Fig.~\ref{FIG_DoubleStressor}(b) is most strongly influenced by the parameters $D$ and $t_\text{stressor}$. 
Figure~\ref{FIG_DoubleStressor}(c) shows a typical map of $\varepsilon_{xy}$ for a stressor aspect ratio of $L$:$W=4$.
Figure~\ref{FIG_DoubleStressor}(d) shows a linecut through this same data, along the dashed line shown in Fig.~\ref{FIG_DoubleStressor}(b), which passes through the center of the device.
We note that the device parameters for these two plots correspond to $L=2000$~nm, $W=500$~nm, $D=250$~nm, $t_\text{stressor}=30$~nm, and $t_\text{SiGe}=30$~nm, and that all data points displaying a blue star in this figures also use these same device parameters.

Figure~\ref{FIG_DoubleStressor}(e) shows that, as $D$ increases, the shear strain initially increases, while reaching a maximum and decaying for larger $D$ values. 
Here, the increasing strain can be understood in terms of constructive interference between the strain patterns, while the decreasing strain is caused by the lateral decay of the strain fields.
For this particular geometry, we find that the strain is largest for the separation range of $D=150$-\SI{350}{\nano\meter}, which happens to be well-suited for integrating the stressors above or below quantum dot plunger gate stacks, without any undo interactions (see below). 
This geometry thus offers improved scalability over some previous strain-engineering proposals, such as etched trenches~\cite{Woods2024CouplingOscillations}, since the stressors here can be positioned close to the quantum dots, without requiring a large stressor footprint or other significant constraints on the gate routing or device layout.

\begin{figure*}[t!]
\begin{center}
\includegraphics[width=0.65\textwidth]{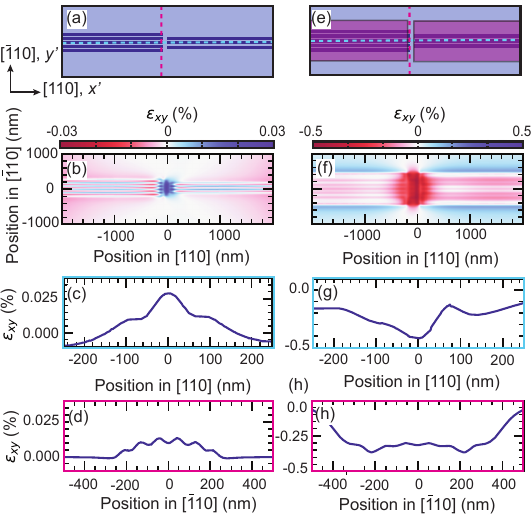}
\end{center}
\vspace{-0.5cm}
\hspace{-1cm}
\caption{Impact of Al wires on the shear strain arising from overlying stressors. 
(a) Top-down schematic of a simplified double-dot gate structure with five Al wires on the dot side (left) and three wires on the charge-sensing side (right). 
Note the absence of any stressors in this geometry.
The wires have a pitch of \SI{80}{\nano\meter}, a width of \SI{60}{\nano\meter}, a thickness of \SI{70}{\nano\meter}, and a depositional stress of $\sigma_{\mathrm{Al}} = $~\SI{60}{\mega\pascal} \cite{Guisbiers2006ResidualSubstrates}. 
The horizontal and vertical dashed lines show the positions of linecuts that will be evaluated below.
(b) A 2D plot of the shear strain $\varepsilon_{xy}$ in the quantum well, induced by the Al wires. 
(c) A linecut of the results in (b), along the dashed-blue line shown in (a). 
(d) A linecut of the results in (b), along the dashed-magenta line shown in (a). 
(e) Top-down schematic of the same metal gate structure shown in (a), with stressors deposited conformally on top of the wires as shown. 
Here, the stressors have a thickness of $t_\text{stressor} = 200$ nm and residual stress $\sigma_{\mathrm{TiN}} = $~\SI{-3.8}{\giga\pascal} \cite{Ohya2014RoomResonators}.
(f) A 2D plot of the shear strain $\varepsilon_{xy}$, in the device shown in (e), including both Al wires and stressors. 
(g) A linecut of the results in (f), along the dashed-blue line shown in (e). 
(h) A linecut of the results in (f), along the dashed-magenta line shown in (e). 
Note the very different scales for the strain in (b-d) and (f-h).
}
\vspace{-1mm}
\label{FIG_DoubleStressorWithWires}
\end{figure*}

Figure~\ref{FIG_DoubleStressor}(f) shows the strain at the pink-starred location, between two stressors, as a function of the aspect ratio $L$:$W$.
The results indicate a similar (modest) improvement in shear strain, as already observed in the single stressor geometry in Fig.~\ref{FIG_SingleStressor}(h). 
While this weak dependence does not lead to significant increases in the shear strain, it does enable greater flexibility in designing structures with realistic gate layouts and routing requirements.

As shown in Fig.~\ref{FIG_DoubleStressor}(g), thicker stressors induce more strain in the quantum well, as expected. 
The strain increases with stressor thickness, and with increasing thicknesses; the rate of increase is sublinear, corresponding to diminishing gains in shear strain with increasing stressor thickness. 

In Fig.~\ref{FIG_DoubleStressor}(h), we investigate the dependence of shear strain on the thickness of the SiGe spacer layer.
As expected, decreasing $t_\text{SiGe}$ causes an increase in $\varepsilon_{xy}$ as the stressor is brought closer to the well. 
However, this dependence is considerably weaker than the dependence on $t_\text{stressor}$ in Fig.~\ref{FIG_DoubleStressor}(g). 
Consequently, $t_\text{SiGe}$ can be adjusted to satisfy heterostructure and fabrication requirements while still maintaining significant tunability via $t_\text{stressor}$.

In practice, the achievable stressor thickness and intrinsic film stress are not fully independent parameters. These parameters are governed in part by the deposition process and growth conditions \cite{Thornton1989STRESS-RELATEDFILMS, Windischmann1991IntrinsicFilms}. Increasing the stressor thickness can modify the intrinsic stress of the film and may introduce fabrication challenges such as cracking, delamination, or stress relaxation \cite{Ohring2001MaterialsEdition}. Consequently, device design requires balancing stressor geometry, film thickness, and achievable residual stress within experimentally realistic fabrication limits. 

\subsection{Stressors integrated with simple gates}
\label{Sec:Stressor_Gates}

While Secs.~\ref{Sec:Single_Stressor} and \ref{Sec:Double_Stressor} show that substantial shear strain can be induced both directly below and near the edges of deposited stressors, realistic quantum dot devices must also incorporate metallic gates to provide electrostatic control. It is therefore important to understand how the presence of these metal gates impacts the strain experienced by the quantum dots. While metal gates in the absence of stressors have been shown to produce strains on the order of several hundredths of a percent \cite{Frink2025ReducingStacking, Corley-Wiciak2023LatticeDevices, Corley-Wiciak2022NanoscaleDevice}, their impact on the strain fields produced by stressors is unclear. 
Moreover, it is known that gate-layer stacking can have varying effects on the total strain \cite{Frink2025ReducingStacking}.

In this subsection, to investigate these questions, we study stressors formed on top of long aluminum wires. In particular, we consider a simple geometry in which the stressors are deposited on a small number of gates, as shown in Fig.~\ref{FIG_DoubleStressorWithWires}. This model provides intuition for how gates underneath the stressors modify the strain fields within the quantum well and provides insight into potential constraints on stressor integration. We find that the strain fields below the gates can be significantly suppressed. Importantly, however, we show that the shear strain can remain large in the edge regions, allowing us to exploit the strategies already explored in Sec.~\ref{Sec:Double_Stressor}.

We first consider a simplified double-dot gate structure with five Al metal wires defined to be on the ``dot side'' of the device (left) and three wires defined to be on the ``charge-sensing side'' (right), with all gates formed directly on top of the oxide layer, as shown in Fig.~\ref{FIG_DoubleStressorWithWires}(a). The gap between the the ends of the dot-side wires and the charge-sensing wires is \SI{110}{\nano\meter}, approximating a typical gap size allotted for a central screening gate. 
The wires have a pitch of \SI{80}{\nano\meter}, a width of \SI{60}{\nano\meter}, a thickness of \SI{70}{\nano\meter}, and a depositional stress of $\sigma_{\mathrm{Al}} =$~\SI{60}{\mega\pascal}~\cite{Guisbiers2006ResidualSubstrates}. 
The wires are encased in a native \SI{2}{\nano\meter} AlOx film. The wires generate local strain fluctuations due to a combination of depositional stress, geometry, edge effects, and thermal contraction, as detailed in Appendix~\ref{appendix:COMSOL}. Figures~\ref{FIG_DoubleStressorWithWires}(b)-\ref{FIG_DoubleStressorWithWires}(d) show simulation results for the shear strain produced by the wires described above, in the absence of an integrated stressor.
In Fig.~\ref{FIG_DoubleStressorWithWires}(c), we report the shear strain along the horizontal dashed-blue linecut shown in Fig.~\ref{FIG_DoubleStressorWithWires}(a).
Here we see the wires induce a relatively low level of shear strain, focused mainly in the gap between the gates. 
In Fig.~\ref{FIG_DoubleStressorWithWires}(d), we show the shear strain along the vertical dashed-magenta linecut shown in Fig.~\ref{FIG_DoubleStressorWithWires}(a), representing the locations of the dots.
Here, the shear strain variations are consistent with those reported in Ref.~\cite{Frink2025ReducingStacking}. The positive sign of the observed strain is a combined effect of the positive depositional stress in the Al and thermal contractions.

\begin{figure}[t]
\begin{center}
\includegraphics[width=0.47\textwidth]{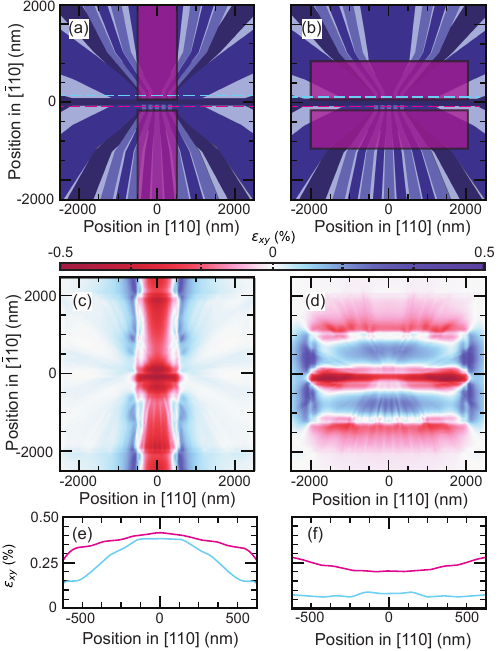}
\end{center}
\vspace{-0.5cm}
\hspace{-1cm}
\caption{Strain simulations of a quadruple quantum dot device with deposited TiN stressors.
(a) Top-down schematic of a quad-dot device with stressors oriented along the $[\bar{1}10]$ lattice direction. The stressors have width $W = 1000$ nm, length $L = 4000$ nm, and thickness $t_\text{stressor} = 200$ nm, with residual stress $\sigma_{\mathrm{TiN}} = -3.8~\mathrm{GPa}$, consistent with Ref.~\cite{Ohya2014RoomResonators}. 
(b) Top-down schematic of the same device with stressors oriented along the $[110]$ lattice direction. The stressor dimensions are identical to (a), but rotated, such that the long and short axes are interchanged ($L \leftrightarrow W$), illustrating a layout compatible with horizontal scaling.
(c) Shear strain $\varepsilon_{xy}$ for the device in (a). 
(d) Shear strain $\varepsilon_{xy}$ for the device in (b). 
(e) Linecuts of $\varepsilon_{xy}$ beneath the quantum dots [pink dashed line in (a)] and beneath the charge-sensing dots [blue dashed line in (a)]. 
(f) Linecuts of $\varepsilon_{xy}$ beneath the quantum dots [pink dashed line in (b)] and beneath the charge-sensing dots [blue dashed line in (b)].}
\vspace{-1mm}
\label{FIG_RealisticDevice}
\end{figure}

Next, we integrate stressors conformally over the tops of the gates, as illustrated in Fig.~\ref{FIG_DoubleStressorWithWires}(b), assuming a stressor thickness of $t_\text{stressor} =$~\SI{200}{\nano\meter} and a depositional stress of $\sigma_{\mathrm{TiN}} =$~\SI{-3.8}{\giga\pascal}, consistent with literature values~\cite{Ohya2014RoomResonators}. 
Close inspection shows that, for the geometry considered here, the metal gates stick out slightly further into the gap than the stressors.
The results of shear-strain simulations are shown in Figs.~\ref{FIG_DoubleStressorWithWires}(f)-\ref{FIG_DoubleStressorWithWires}(h).
In Fig.~\ref{FIG_DoubleStressorWithWires}(f), the presence of the Al wires under the stressor is clearly seen to modulate the strain below the stressor. 
However, the magnitude of the shear strain in the gap is strongly amplified by the presence of the stressors, as seen in Fig.~\ref{FIG_DoubleStressorWithWires}(g),  along the horizontal dashed-blue linecut illustrated in Fig.~\ref{FIG_DoubleStressorWithWires}(e).
In Fig.~\ref{FIG_DoubleStressorWithWires}(h), along the vertical dashed-magenta linecut, where the dots are located, the overall magnitude of the shear strain is greatly enhanced compared to Fig.~\ref{FIG_DoubleStressorWithWires}(d), where no stressor was present. 
Although the wires still cause small fluctuations of the shear strain, the stressor-induced strain is much larger than the strain generated by the gates alone. 
However, since the shear-strain enhancement from the stressors is largest near the gap, it is important to place the dots near the end of the stressor; the region under the bulk of the stressor would be a relatively poor location for centering the dots.

When the stressors are deposited conformally on the gates, the underlying metal wires lift portions of the stressors away from the quantum well. 
Thus, the strain patterns in this system are complicated by two competing factors: stress caused by the gates (including the depositional stress in the wires and the strain due to thermal contraction) and the greater separation between the stressor and the quantum well. Since the strain induced by the stressor decreases with distance from the well [as discussed previously in Fig.~\ref{FIG_DoubleStressor}(h), where we considered changes in the thickness of the SiGe spacer layer], the presence of metal gates similarly moves the stressors farther from the quantum well, thereby reducing the strain magnitude experienced in the quantum well. 
This effect can be compensated, in part, by increasing the stressor thickness, as shown in Appendix~\ref{Sec:Thickness}.
Here, we find that achieving strain magnitudes similar to Fig.~\ref{FIG_DoubleStressor}, where the stressors were formed directly on the heterostructure surface, requires stressor thicknesses approximately twice that of the underlying wires.

\subsection{Stressors integrated with a realistic device}
\label{Sec:Stressor_Device}

In this subsection, we study the effect of stressors formed on top of a realistic gate design for quantum dot qubits. 
Specifically, we consider a quadruple quantum dot design similar to Ref.~\cite{Neyens2019MeasurementsArray}, as illustrated in Figs.~\ref{FIG_RealisticDevice}(a) and \ref{FIG_RealisticDevice}(b). 
In the lower half of the device, the plunger gates define four quantum dots and two 2DEG reservoirs, one each on either side. 
In the upper half of the device, the plungers define two charge sensing dots with three 2DEG reservoirs. 
Stressors are formed over the gates and centered over the dot channels.

Two orientations of stressors are considered, the first where the short edge of the stressor is aligned with the dot channel [Figs.~\ref{FIG_RealisticDevice}(a,c,e)], and the second where the long edge of the stressor is aligned with the dot channel [Figs.~\ref{FIG_RealisticDevice}(b,d,f)]. 
The simulations assume the same heterostructure and device parameters as Fig.~\ref{FIG_DoubleStressorWithWires}, except for the presence of \emph{three} overlapping gate layers. 
In Fig.~\ref{FIG_RealisticDevice}, the gate layers are shown in different shades of blue (dark to light), representing screening gates, plunger and reservoir gates, and tunnel-barrier gates, respectively.
(See \cite{Neyens2019MeasurementsArray} for further details.)
Each gate layer is surrounded by a thin layer of oxide, which can have nontrivial effects on the strain patterns in the quantum well \cite{Frink2025ReducingStacking}.

Similar to results described in previous subsections, the presence of metal gates beneath the stressors is manifested as small strain variations, as seen in Figs.~\ref{FIG_RealisticDevice}(c) and \ref{FIG_RealisticDevice}(d), for the two different stressor orientations. 
In Figs.~\ref{FIG_RealisticDevice}(e) and \ref{FIG_RealisticDevice}(f) we plot linecuts through the two devices, evaluated along the horizontal lines shown in Figs.~\ref{FIG_RealisticDevice}(a) and \ref{FIG_RealisticDevice}(b), corresponding to the dot channel (dashed magenta) and the charge-sensor channel (dashed blue).
As we found in earlier simulations, the strain induced near the short edge of the stressor is larger than that near the long edge. However, for the configuration shown in  Fig.~\ref{FIG_RealisticDevice}(b), the shear strain is relatively large and uniform across the entire dot channel, allowing strain to be applied to a long, linear array of quantum dots \cite{George202512-Spin-QubitLine, MembersoftheHRLQuantumTeam2026AUnit}. 

\section{Discussion}

Several approaches have been proposed for increasing valley splitting in Si/SiGe quantum dots. Increasing the Ge content of the quantum well enhances the average valley splitting, but because this enhancement arises from random-alloy disorder, spatial variations still produce regions of vanishingly small valley splitting \cite{Marcks2025ValleyGermanium, Losert2024StrategiesDisorder}. 
In contrast, deterministic approaches based on engineered heterostructures --- such as short-period Wiggle Wells and ultra-sharp Si/SiGe interfaces --- can also enhance the valley splitting but place stringent demands on the epitaxial growth precision \cite{Losert2024StrategiesDisorder}. 
As an alternative approach for deterministically enhancing the valley splitting, the long-period Wiggle Well has been proposed in combination with trenches etched along the $[110]$ and $[\bar{1}10]$ crystallographic directions, with quantum dots positioned between the trenches to take advantage of the resulting shear strain~\cite{Woods2024CouplingOscillations}. Although this approach can generate the required symmetry-breaking strain, the trenches occupy substantial area and constrain where the gate-defined devices can be placed. This becomes increasingly restrictive as quantum dot arrays are scaled to larger dimensions.

In this work, we have shown that stressors deposited on the surface of a gated quantum dot device formed in a Wiggle Well heterostructure can generate sufficient shear strain to deterministically enhance the valley splitting, even in the presence of a realistic gate stack. It has been shown previously that anisotropy in strain between $[110]$ and $[\bar{1}10]$ can be used to generate shear strain $\varepsilon_{xy}$ in the quantum well and, when combined with an oscillating Ge concentration, provide the symmetry breaking required to couple valleys and deterministically enhance the valley splitting~\cite{Woods2024CouplingOscillations}. 
Here, we build upon that result by investigating whether deposited stressors can generate the required shear strain in realistic gated quantum dot devices. 

Deposited stressors provide greater flexibility than etched trenches because they can be integrated above the gate architecture rather than requiring large regions of the heterostructure to be removed. Moreover, the stressors extend along the full length of a quantum dot array, allowing many dots to be formed between a single pair of stressors without requiring repeated stressor structures, as shown in Fig.~\ref{FIG_RealisticDevice}(b). There are, however, important limitations. The region of large shear strain between neighboring stressors is relatively narrow, and the magnitude of $\varepsilon_{xy}$ decreases as the stressor separation is increased, as shown in Fig.~\ref{FIG_DoubleStressor}(e). 
The largest and most spatially uniform shear strain occurs in the region directly between the two stressors and decreases toward the stressor edges, limiting the width over which a large and relatively uniform $\varepsilon_{xy}$ can be maintained, as shown in Figs.~\ref{FIG_DoubleStressor}(c) and \ref{FIG_DoubleStressor}(d). 
The geometries considered here are therefore naturally compatible with linear and relatively narrow multi-row arrays (e.g., $1\times N$, $2\times N$, and $3\times N$), while substantially wider 2D arrays would require additional stressors or modified stressor geometries to maintain large shear strain across the full device. 
This limitation is less restrictive for near-term architectures in which quantum dot arrays remain only a few rows wide, but stressor layouts will need to evolve alongside larger-scale device designs.

Surface stressors can generate the desired shear strain levels when patterned anisotropically along the $[110]$ and $[\bar{1}10]$ crystallographic directions. The depositional stress in the deposited film drives elastic deformation of the underlying heterostructure, while the stressor geometry breaks the in-plane symmetry between the two crystallographic directions. The resulting deformation therefore contains a substantial $\varepsilon_{xy}$ component in the quantum well. Consequently, both the crystallographic orientation and the geometry of the stressors are important design parameters for maximizing the shear strain experienced by the quantum dots.

The eventual question of this work is not whether it is possible to generate the desired shear strain, but whether sufficiently large shear strains can be maintained after incorporating the stressors into a realistic gate stack. 
Although the inclusion of metallic gates modifies the strain distribution in the quantum well, sufficiently thick stressors can still generate shear strains well above the $\varepsilon_{xy}>0.1\%$ threshold required for substantial deterministic enhancement of the valley splitting. The gates affect the strain environment in two ways: they exert their own stress on the heterostructure, and they increase the physical separation between the stressor and the quantum well by lifting portions of the stressor away from the heterostructure surface. 
Both effects can reduce the magnitude of shear strain transferred to the quantum well region. 
However, as shown in Appendix~C and Fig.~\ref{FIG_Thickness}, this reduction can be compensated by increasing the stressor thickness. 
It is important to note that the depositional stress in a stressor film is not independent of its thickness; the stress magnitude, and even its sign, can vary with deposition conditions and film thickness, so the stressor material and deposition process must be co-optimized to achieve the desired stress at the required thickness~\cite{Huff2022Citation:Manufacturing, Ohring2001MaterialsEdition}.

Electrical isolation of the stressors from the gate stack provides an additional fabrication consideration. A dielectric layer is required between the stressors and the metallic gates to prevent electrical shorting. While such a layer enables integration of the stressor directly above the gate structure, it also introduces an additional material interface near the quantum well that could contribute to charge noise. The choice and thickness of this dielectric therefore represent a tradeoff between electrical isolation, strain transfer, and the electrostatic environment of the quantum dots. Future experimental devices will need to determine whether this additional interface produces a measurable degradation in charge-noise performance and whether the stressor and dielectric stack can be optimized to minimize this effect. Nonetheless, the ability to deposit the stressor above the completed gate stack offers a significant advantage in that the stressor process can be optimized independently of the gate fabrication.

In summary, we have shown that patterned stressors, aligned along the $[110]$ and $[\bar{1}10]$ crystallographic directions and deposited over gates, are capable of producing sufficient shear strain in a Wiggle Well heterostructure to enable deterministic enhancement of the valley splitting energy. The predicted shear strains $(\varepsilon_{xy} \geq 0.1\%)$ are large enough to generate a deterministic valley splitting contribution exceeding $100~\mu\mathrm{eV}$ for a long-period Wiggle Well with an average Ge concentration of 4.5\%, consistent with heterostructures that have been experimentally demonstrated~\cite{Woods2024CouplingOscillations, McJunkin:2022p7777}. Compared to alternative approaches for enhancing the valley splitting, including etched trenches, ultra-sharp interfaces, and short-period Wiggle Wells, stressors deposited over quantum-dot gates offer a potentially more fabrication-compatible and scalable route for valley splitting engineering while maintaining compatibility with realistic gate architectures. These results establish deposited stressors combined with a Wiggle Well heterostructure as a promising route for deterministic control of the valley splitting in future Si/SiGe spin-qubit devices. 

\section*{Acknowledgments}
This work was sponsored in part by the Army Research Office under Award No.\ W911NF-23-1-0110. The views, conclusions, and recommendations contained in this document are those of the authors and are not necessarily endorsed by nor should they be interpreted as representing the official policies, either expressed or implied, of the Army Research Office or the U.S. Government. The U.S.\ Government is authorized to reproduce and distribute reprints for U.S.\ Government purposes notwithstanding any copyright notation herein.  Sandia National Laboratories is a multimission laboratory managed and operated by National Technology and Engineering Solutions of Sandia, LLC, a wholly owned subsidiary of Honeywell International Inc., for the U.S. Department of Energy’s National Nuclear Security Administration under contract DE-NA0003525.

E.S.J., B.D.W., M.F., and M.A.E.\ have applied for a patent on strained Si/SiGe quantum wells. 
The remaining authors declare no competing interests.

\appendix

\section{Simulation methods}
\label{appendix:COMSOL}
Strain calculations were performed using the Solid Mechanics module in COMSOL Multiphysics~\cite{COMSOLABV.5.6}. These simulations require specification of the modulus of elasticity $E$, Poisson's ratio $\nu$, and coefficients of thermal expansion (CTE) $\alpha$ for each material. While all three parameters are temperature dependent, the elastic moduli and Poisson's ratios exhibit only weak temperature dependence~\cite{Hopcroft2010WhatSilicon}; therefore, room-temperature values are used for these quantities, and unless otherwise specified, these values are adopted from the COMSOL parameter library. To account for the temperature dependence of the CTEs, we define an effective temperature-averaged coefficient $\bar{\alpha} = (T_i - T_f)^{-1}\int_{T_f}^{T_i} \alpha(T)\, dT$, where $T_i$ and $T_f$ are the initial and final temperatures, respectively. The integrals are approximated using temperature-dependent CTE data from Refs.~\cite{Bradley2013PropertiesTemperatures, Simon1994CRYOGENICMAGNETS, McSkimin1953MeasurementWaves,Levinshtein2001PropertiesSiGe, Corruccini1961ThermalLiterature,Burke1962ThermalC, Touloukian1977ThermalSolids}. For the SiGe alloy, the CTE is evaluated by combining the Si and Ge data using Vegard's law. The resulting effective CTEs are: $\alpha_{\text{Si}} = 0.76 \times 10^{-6}\,\text{K}^{-1}, 
\alpha_{\text{SiGe}} = 1.52 \times 10^{-6}\,\text{K}^{-1}, 
\alpha_{\text{Al}_2\text{O}_3} = 3.3 \times 10^{-6}\,\text{K}^{-1}, 
\alpha_{\text{TiN}} = 4.4 \times 10^{-6}\,\text{K}^{-1}, 
\alpha_{\text{Al}} = 14.6 \times 10^{-6}\,\text{K}^{-1},\ 
\text{and }\alpha_{\text{Si}_3\text{N}_4} = 0.5 \times 10^{-6}\,\text{K}^{-1}$. 

Because Si and SiGe are elastically anisotropic, we include the full compliance tensors for both materials in the COMSOL simulations \cite{Frink2025ReducingStacking}. Biaxial strain in the quantum well, which arises from lattice mismatch between the Si quantum well and the strain-relaxed SiG, is given by $\varepsilon_{xx} = \varepsilon_{yy} = 1.2707\%$ and $\varepsilon_{zz} = -0.9798\%$. This strain is implemented in the simulations via initial-strain parameters. 

Depositional stress in the metal gates is incorporated through initial-stress parameters, which include the effects of thermal contraction as the metal is cooled from deposition temperatures to $293~\mathrm{K}$. For Al wires and gates, we assume a depositional stress of $60~\mathrm{MPa}$~\cite{Guisbiers2006ResidualSubstrates}. For TiN stressors we adopt the values from two literature sources, $-2.6 ~\mathrm{GPa}$ for a 30 nm thick film~\cite{Corley-Wiciak2022NanoscaleDevice} and $-3.8 ~\mathrm{GPa}$ for a 200 nm thick film~\cite{Ohya2014RoomResonators}. 
The elastic modulus for TiN, $E_\mathrm{{TiN}}=106~\mathrm{GPa}$, is adopted from the same literature value as the 30~nm thick film~\cite{Corley-Wiciak2022NanoscaleDevice}.
No depositional stress is assumed for Al$_2$O$_3$, owing to the wide range of reported values, including sign changes, and their sensitivity to deposition conditions and thermal post-processing \cite{Ylivaara2022ThermomechanicalDeposition}. 
This approximation is not expected to affect the qualitative conclusions, but may introduce some quantitative deviations.

\begin{figure}[t]
\begin{center}
\includegraphics[width=0.47\textwidth]{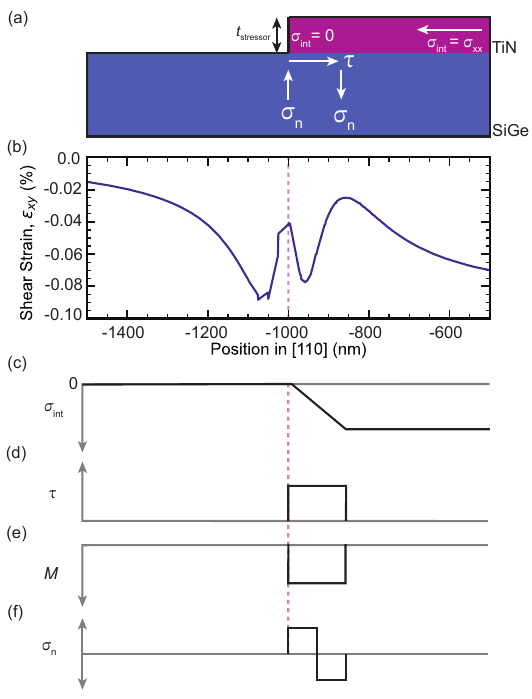}
\end{center}
\vspace{-0.5cm}
\hspace{-1cm}
\caption{Qualitative illustration of the stress-transfer mechanism responsible for the ``double-dip'' edge feature. (See Refs.~\cite{Hibbeler2014MechanicsMaterials, Ohring2001MaterialsEdition, Freund2004ThinEvolution} for additional details.)
(a) Cross-sectional schematic of a long deposited stressor on a SiGe surface.
(b) Simulated shear strain $\varepsilon_{xy}$ beneath the stressor showing the localized edge effect, reproduced frm Fig.~\ref{FIG_SingleStressor}(d).
(c)-(f) Qualitative diagrams illustrating (c) the relaxation of the internal stress of the film, (d) the resulting interfacial shear stress, (e) the associated bending moment, and (f) the normal stresses required to maintain equilibrium.
}
\vspace{-1mm}
\label{FIG_EdgeEffect}
\end{figure}

Free boundary conditions are applied on all surfaces except the bottom surface, where a constraint is imposed to match the bulk thermal contraction of Si, thereby emulating a semi-infinite substrate~\cite{COMSOLABV.5.6}. 
When computing shear strain, we use the strain evaluated in the 2DEG plane, which is assumed to lie $1.5~\text{nm}$ below the top interface of the quantum well, where the low-energy electronic wave functions of the quantum well are concentrated. The shear strain is computed using Eq.~(\ref{eq:shear_transform}), which maps the strain from the $[110], [\bar{1}10], [001]$ device frame to the $[100],[010],[001]$ crystallographic frame. 

For the realistic quadruple-dot device simulations of Sec.~\ref{Sec:Stressor_Device}, the stressor was modeled as a uniform 200~nm film deposited on top of an overlapping gate structure, with a flat top, due to technical challenges with constructing this 3D model in COMSOL. This approximation underestimates the stressor volume above the gates and therefore provides a conservative estimate of the strain transferred to the quantum well while minimizing the computational cost.

\section{Edge effects at the stressor boundary}
\label{Sec:EdgeEffect}

The strain distributions presented in the main text exhibit ``double-dip'' features near the edges of the deposited stressors. These features are not numerical artifacts but arise naturally from the transfer of stress between a stressed film and the underlying substrate. The qualitative mechanics of this process are well established in the mechanics of materials and thin-film literature~\cite{Ohring2001MaterialsEdition, Freund2004ThinEvolution}. 
Figure~\ref{FIG_EdgeEffect} provides a schematic illustration of the mechanism responsible for these features.

Consider a thin film, with an internal (residual) stress $\sigma_{\mathrm{int}}$, bonded to the surface of an elastic substrate [Fig.~\ref{FIG_EdgeEffect}(a)]. 
Far from a free edge, the film carries an approximately uniform internal stress. At the edge of the film, however, the stress must relax because no external traction exists beyond the boundary. Consequently, the internal stress varies rapidly near the edge of the film, as illustrated schematically in Fig.~\ref{FIG_EdgeEffect}(c). 

\begin{figure}[t]
\begin{center}
\includegraphics[width=0.47\textwidth]{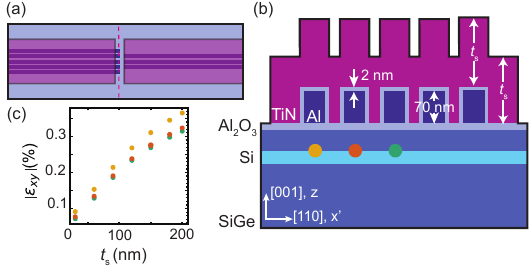}
\end{center}
\vspace{-0.5cm}
\hspace{-1cm}
\caption{Strain dependence on stressor thickness, for a stressor fabricated on top of the gate electrodes.
(a) Top-down schematic of a simplified double-dot gate structure shown in Fig.~\ref{FIG_DoubleStressorWithWires}(e), with five Al wires on the dot side (left) and three on the charge-sensing side (right), and with stressors deposited conformally on top of the wires and centered above the dots. 
The device parameters are the same as described in Fig.~\ref{FIG_DoubleStressorWithWires}.
(b) Cross-sectional schematic of the dot side of the device depicted in (a), with the thickness of the oxide and the height of the wires indicated. 
The thickness of the stressor is indicated as $t_s=t_\text{stressor}$. 
The shear strain is evaluated in the quantum well under the center of three gates, as indicated by yellow, orange, and green circles, along the dashed line shown in (a).
(c) Simulated shear strain results in the quantum well under the center of three gates, as indicated in (b), using the same color scheme. 
}
\vspace{-1mm}
\label{FIG_Thickness}
\end{figure}

Any spatial gradient in the film stress must be balanced by an interfacial shear stress that transfers the load from the film into the substrate. Force equilibrium on a differential element of the film therefore requires that \cite{Ohring2001MaterialsEdition}
\begin{equation}
    \frac{d\sigma_\text{int}}{dx}=-\frac{\tau}{t_\text{stressor}}
\end{equation}
where $\tau$ is the interfacial shear stress. Consequently, a localized shear stress develops near the stressor edge [Figs.~\ref{FIG_EdgeEffect}(a) and \ref{FIG_EdgeEffect}(d)], transferring the load from the film to the substrate. 

\begin{figure}[t]
\begin{center}
\includegraphics[width=0.47\textwidth]{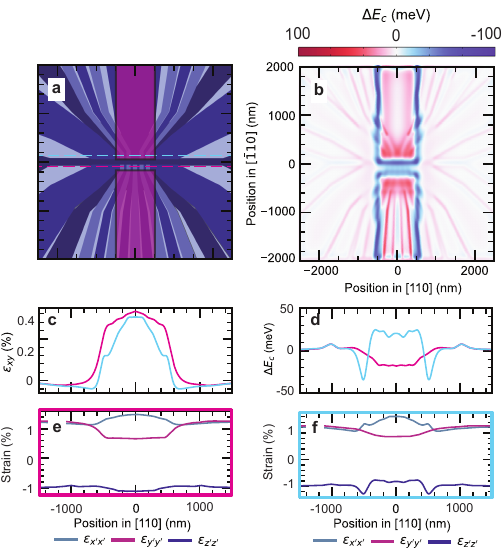}
\end{center}
\vspace{-0.5cm}
\hspace{-1cm}
\caption{Conduction-band potential landscapes, for a quadruple quantum dot device with deposited TiN stressors.
(a) Top-down schematic of the quad-dot device with stressors, from Fig.~\ref{FIG_RealisticDevice}(a). 
(b) Conduction band offset $\Delta E_c$, based on Eq.~(\ref{DeltaEc}), for the device in (a). 
(c) Linecuts of the shear strain $\varepsilon_{xy}$ beneath the quantum dots [pink line in (a)] and beneath the charge sensing dots [blue line in (a)].
(d) Corresponding conduction-band offsets beneath the quantum dots and sensing dots.
(e,f) $\varepsilon_{x'x'}$ (light blue), $\varepsilon_{y'y'}$ (magenta), and $\varepsilon_{z'z'}$ (dark blue) components of the strain in the well, evaluated below the quantum dots [pink line in (a)] and the charge-sensing dots [blue line in (a)].
}
\vspace{-1mm}
\label{FIG_Energy}
\end{figure}

\begin{table*}[t]
\small
\centering
\begin{tabular}{cccccc}
\toprule
\textbf{Reference} & \textbf{Deposition type} & \textbf{Thickness (nm)} & \textbf{Stress (GPa)} & \textbf{\% Strain} & \textbf{$E_\text{VS}$ ($\mu$eV)} \\
\midrule
\multicolumn{6}{c}{\textbf{TiN}} \\
\midrule
Ohya \cite{Ohya2014RoomResonators} & Magnetron sputter & 200 & $-3.8$ & $-0.556$ & $>500$\\
Corley-Wiciak \cite{Corley-Wiciak2023LatticeDevices} & Sputter & 30 & $-1.6$ & $-0.076$ & $\approx100$\\
Corley-Wiciak \cite{Corley-Wiciak2023LatticeDevices} & Magnetron sputter & 30 & $-2.6$ & $-0.128$ & $\approx200$\\
Hussain \cite{Hussain2006DepositionCharacteristics} & PVD & 10 & $-6.32$ & $-0.127$ & $\approx200$ \\
Hussain \cite{Hussain2006DepositionCharacteristics} & CVD & 10 & $-1.37$ & $-0.025$ & $\approx25$\\
Hussain \cite{Hussain2006DepositionCharacteristics} & ALD & 10 & $-2.38$ & $-0.046$ & $\approx75$ \\
\midrule
\multicolumn{6}{c}{\textbf{Si$_3$N$_4$}} \\
\midrule
Irene \cite{Irene1976ResidualFilms} & CVD & 200 & $+1.2$ & $+0.151$ & $\approx225$ \\
Shi \cite{Shi2006ModifyingImplantation} & LPCVD & 200 & $-1.2$ & $-0.165$ & $\approx250$ \\
Capellini \cite{Capellini2014TensileProcess} & PECVD & 400 & $-2.5$ & $-0.412$ & $>500$ \\
\bottomrule
\end{tabular}
\caption{Calculated shear strain and deterministic valley splittings, for TiN and Si$_3$N$_4$ thin stressor films applied using a range of depositional techniques.}
\label{tab:reported_stressors}
\end{table*}

The transferred load generates a localized bending moment $M$ at the film-substrate interface, because the internal force carried by the film and the interfacial shear force act at different heights within the structure [Fig.~\ref{FIG_EdgeEffect}(e)]. To satisfy mechanical equilibrium, additional normal stresses $\sigma_n$ develop in the film and substrate near the edge of the stressor [Fig.~\ref{FIG_EdgeEffect}(f)]. 
These localized stresses perturb the strain field in the substrate and give rise to the characteristic dip-peak feature observed in the simulated strain profile shown in Fig.~\ref{FIG_EdgeEffect}(b). 

Although the detailed stress distribution depends on the elastic properties and geometry of the film-substrate system, the underlying mechanism is generic: relaxation of the residual film stress at a free edge produces localized interfacial shear stresses that transfer the load into the substrate, resulting in a concentrated stress field, with a profile like that observed in our COMSOL simulations [Fig.~\ref{FIG_EdgeEffect}(b)] \cite{Ohring2001MaterialsEdition, Freund2004ThinEvolution}.

\section{Strain dependence on stressor thickness}
\label{Sec:Thickness}

In Sec.~\ref{Sec:Stressor_Gates}, it was shown that integrating stressors on top of gate electrodes is feasible, but reduces the strain transferred to the quantum well. This reduction arises because the gate stack increases the effective separation between the stressors and the quantum well, weakening the strain transfer. To determine the stressor thickness required to recover the desired strain levels, we simulated the geometry shown in Fig.~\ref{FIG_DoubleStressorWithWires}.

To study this behavior in more detail, we present additional simulation results in Fig.~\ref{FIG_Thickness}, obtained by sweeping the thickness of the stressor.
In Fig.~\ref{FIG_Thickness}(a), we again consider the simple straight-gate geometry, with a double-stressor arrangement, from Sec.~\ref{Sec:Stressor_Gates}.
Because the induced strain varies across the gate array, we evaluate the shear strain in the quantum well at three representative locations: beneath the outermost gate, beneath the adjacent inner gate, and beneath the central gate, as indicated by circular dots in the cross-sectional view shown in Fig.~\ref{FIG_Thickness}(b). The stressor thickness $t_s=t_\text{stressor}$ is then varied to determine how the transferred shear strain depends on film thickness. 

Figure~\ref{FIG_Thickness}(c) shows $t_\text{stressor}$ sweep results, indicating that the transferred shear strain increases monotonically with stressor thickness at all three evaluation locations. 
We observe a slight decrease in the rate of increase for larger thicknesses. 
The outermost gate experiences the largest shear strain because it lies closest to the stressor edge, while the two interior locations exhibit nearly identical strain values. Although the gate stack reduces the transferred strain, as compared with the no-gate geometry, these results show that thicker stressors can at least partially compensate for this effect.
These results indicate that stressor integration above gate electrodes remains a viable approach for generating large shear strains in the quantum well, provided sufficiently thick stressor films are deposited.

\section{Other Stressor Materials}
\label{Sec:OtherMaterials}

Residual stress within a thin film is highly dependent on deposition method and process conditions~\cite{Ohring2001MaterialsEdition}. In this Appendix, the same 4:1 stressor geometry described in Sec.~\ref{Sec:Double_Stressor} is simulated using literature-reported values of film thickness and residual stress for a variety of materials and deposition processes.
The results are reported in Table~\ref{tab:reported_stressors}.
This analysis illustrates the tunability of the strain field and the corresponding valley splitting that can be achieved through the choice of stressor material and deposition conditions. 

In addition to TiN, $\mathrm{Si_3N_4}$ was selected as a representative stressor material because silicon nitride films are widely used for strain engineering in semiconductor devices and can be deposited with either tensile or compressive residual stress depending on the deposition conditions~\cite{Irene1976ResidualFilms, Shi2006ModifyingImplantation}. Silicon nitride stressors have been extensively employed in CMOS strain-engineering schemes, where tensile and compressive $\mathrm{Si_3N_4}$ films are used to transfer strain into transistor channels and thereby enhance carrier mobility and device performance~\cite{ItoMechanicalDesign, ShimizuLocalEnhancement}.

\section{Conduction-band energy variations}
\label{Sec:Energy}

The calculated strain maps, reported in several figures throughout this work, can be converted to conduction-band energy shifts, $\Delta E_c$, through the relation~\cite{Thorbeck2015FormationNanostructures}
\begin{equation}
    \Delta E_c = E_0 + \Xi_u \varepsilon_{zz} + \Xi_d (\varepsilon_{xx} + \varepsilon_{yy} + \varepsilon_{zz}), \label{DeltaEc}
\end{equation}
where $\Xi_u=10.5$ eV and $\Xi_d=1.1$ eV are the deformation potential coefficients~\cite{Fischetti1996BandAlloys}. 
Here, the reference energy $E_0$ is chosen to ensure that $\Delta E_c$ approaches zero far away from the active region containing the gates and stressors. 

For design considerations, it is important to note that, while the deposited TiN stressors are designed to generate large shear strains that enhance the valley splitting, they can also affect the local conduction-band offset, which can result in competing outcomes. 
Fig.~\ref{FIG_Energy} shows the calculated conduction-band shift for the realistic quadruple-dot device introduced in Fig~\ref{FIG_RealisticDevice}. Figs.~\ref{FIG_Energy}(a) and~\ref{FIG_Energy}(b) show the device geometry and the resulting conduction-band potential landscape, respectively. Fig.~\ref{FIG_Energy}(c) shows the shear strain, $\varepsilon_{xy}$, along linecuts beneath the quantum-dot channel (pink) and charge-sensing channel (blue), while Fig.~\ref{FIG_Energy}(d) shows the conduction-band shift, $\Delta E_c$.
The spatial variation of the conduction-band shift differs from that of the shear strain because $\Delta E_c$ is dominated by the out-of-plane strain component, $\varepsilon_{zz}$, whereas the quantity of interest for valley-splitting enhancement is the shear strain, $\varepsilon_{xy}$. To illustrate this, Figs.~\ref{FIG_Energy}(e) and (f) show the individual strain components, $\varepsilon_{x'x'}$, $\varepsilon_{y'y'}$, and $\varepsilon_{zz}$, along the quantum-dot and charge-sensing channels, respectively.

At the quantum dot locations, the stressors produce a maximum conduction-band shift of approximately 18~meV. This is nearly an order of magnitude larger than the shift predicted for a comparable device without deposited stressors~\cite{Frink2025ReducingStacking}, consistent with the much larger intrinsic stress of the TiN film (GPa), compared with the aluminum gate electrodes (tens of MPa). 

Although significant, this shift can be compensated electrostatically during device tuning. For a typical gate lever arm of $\alpha \approx 0.1$~meV/mV, an 18~meV conduction-band shift corresponds to a gate-voltage adjustment of approximately $18~\mathrm{meV}/\alpha \approx 180~\mathrm{mV}$. This suggests that the stressor-induced potential does not present a fundamental obstacle to device operation, although it should be considered during gate-stack design and electrostatic tuning. In particular, extending the stressors to overlap the electron-reservoir gates could reduce the relative conduction-band offset between the reservoirs and quantum dots, simplifying electron loading while preserving the desired strain profile.

\bibliography{references}

@misc{COMSOLABV.5.6,
    title = {{ COMSOL Multiphysics{\textregistered} v.\ 5.6}},
    author = {{COMSOL AB}},
    address = {Stockholm, Sweden}
}

@article{George202512-Spin-QubitLine,
    title = {{12-Spin-Qubit Arrays Fabricated on a 300 mm Semiconductor Manufacturing Line}},
    year = {2025},
    journal = {Nano Letters},
    author = {George, Hubert C. and M{\c{a}}dzik, Mateusz T. and Henry, Eric M. and Wagner, Andrew J. and Islam, Mohammad M. and Borjans, Felix and Connors, Elliot J. and Corrigan, J. and Curry, Matthew and Harper, Michael K. and Keith, Daniel and Lampert, Lester and Luthi, Florian and Mohiyaddin, Fahd A. and Murcia, Sandra and Nair, Rohit and Nahm, Rambert and Nethwewala, Aditi and Neyens, Samuel and Patra, Bishnu and Raharjo, Roy D. and Rogan, Carly and Savytskyy, Rostyslav and Watson, Thomas F. and Ziegler, Josh and Zietz, Otto K. and Pellerano, Stefano and Pillarisetty, Ravi and Bishop, Nathaniel C. and Bojarski, Stephanie A. and Roberts, Jeanette and Clarke, James S.},
    number = {2},
    month = {1},
    pages = {793--799},
    volume = {25},
    publisher = {American Chemical Society},
    doi = {10.1021/acs.nanolett.4c05205},
    issn = {15306992},
    pmid = {39721970},
    arxivId = {2410.16583}
}

@article{MembersoftheHRLQuantumTeam2026AUnit,
  title   = {A digitally controlled silicon quantum processing unit},
  author  = {{Members of the HRL Quantum Team and Collaborators}},
  journal = {Nature},
  volume  = {655},
  number  = {8125},
  pages   = {1154--1159},
  month   = jul,
  doi = {10.1038/s41586-026-10754-7},
  year    = {2026}
}

@article{Fischetti1996BandAlloys,
    title = {{Band structure, deformation potentials, and carrier mobility in strained Si, Ge, and SiGe alloys}},
    year = {1996},
    journal = {Journal of Applied Physics},
    author = {Fischetti, M V and Laux, S E},
    number = {4},
    month = {4},
    pages = {2234--2252},
    volume = {80},
    doi = {10.1063/1.363052},
    issn = {0021-8979}
}

@article{Huff2022Citation:Manufacturing,
AUTHOR = {Huff, Michael},
TITLE = {Review Paper: Residual Stresses in Deposited Thin-Film Material Layers for Micro- and Nano-Systems Manufacturing},
JOURNAL = {Micromachines},
VOLUME = {13},
YEAR = {2022},
NUMBER = {12},
pages = {2084},
ARTICLE-NUMBER = {2084},
URL = {https://www.mdpi.com/2072-666X/13/12/2084},
PubMedID = {36557383},
ISSN = {2072-666X},
DOI = {10.3390/mi13122084}
}

@article{Friesen2007_2,
  title = {Valley splitting theory of {SiGe/Si/SiGe} quantum wells},
  author = {Friesen, Mark and Chutia, Sucismita and Tahan, Charles and Coppersmith, S. N.},
  journal = {Phys. Rev. B},
  volume = {75},
  issue = {11},
  pages = {115318},
  numpages = {12},
  year = {2007},
  month = {Mar},
  publisher = {American Physical Society},
  doi = {10.1103/PhysRevB.75.115318},
  url = {https://link.aps.org/doi/10.1103/PhysRevB.75.115318}
}

@article{Woods2024CouplingOscillations,
    title = {{Coupling conduction-band valleys in SiGe heterostructures via shear strain and Ge concentration oscillations}},
    year = {2024},
    journal = {npj Quantum Information},
    author = {Woods, Benjamin D. and Soomro, Hudaiba and Joseph, E. S. and Frink, Collin C.D. and Joynt, Robert and Eriksson, M. A. and Friesen, Mark},
    number = {1},
    pages = {54},
    month = {12},
    volume = {10},
    publisher = {Nature Research},
    doi = {10.1038/s41534-024-00853-6},
    issn = {20566387}
}

@book{Simon1994CRYOGENICMAGNETS,
title = {Cryogenic Properties of Inorganic Insulation Materials for ITER Magnets: A Review},
author = {Simon, N. J.},
doi = {10.2172/761710},
url = {https://www.osti.gov/biblio/761710}, 
place = {United States},
year = {1994},
month = {12}
}

@article{Hussain2006DepositionCharacteristics,
    title = {{Deposition method-induced stress effect on ultrathin titanium nitride etch characteristics}},
    year = {2006},
    journal = {Electrochemical and Solid-State Letters},
    author = {Hussain, M. M. and Quevedo-Lopez, M. A. and Alshareef, H. N. and Larison, D. and Mathur, K. and Gnade, B. E.},
    number = {12},
    pages = {361--363},
    volume = {9},
    doi = {10.1149/1.2359100},
    issn = {10990062}
}

@article{Park2016Electrode-stress-inducedDevices,
    title = {{Electrode-stress-induced nanoscale disorder in Si quantum electronic devices}},
    year = {2016},
    journal = {APL Materials},
    author = {Park, J. and Ahn, Y. and Tilka, J. A. and Sampson, K. C. and Savage, D. E. and Prance, J. R. and Simmons, C. B. and Lagally, M. G. and Coppersmith, S. N. and Eriksson, M. A. and Holt, M. V. and Evans, P. G.},
    number = {6},
    pages = {066102},
    month = {6},
    volume = {4},
    publisher = {American Institute of Physics Inc.},
    doi = {10.1063/1.4954054},
    issn = {2166532X}
}

@article{Gradwohl2025EnhancedSegregation,
    title = {{Enhanced Nanoscale Ge Concentration Oscillations in Si/SiGe Quantum Well through Controlled Segregation}},
    year = {2025},
    journal = {Nano Letters},
    author = {Gradwohl, Kevin P. and Cvitkovich, Lukas and Lu, Chen Hsun and Koelling, Sebastian and Oezkent, Maximilian and Liu, Yujia and Waldh{\"{o}}r, Dominic and Grasser, Tibor and Niquet, Yann Michel and Albrecht, Martin and Richter, Carsten and Moutanabbir, Oussama and Martin, Jens},
    number = {11},
    month = {3},
    pages = {4204--4210},
    volume = {25},
    publisher = {American Chemical Society},
    doi = {10.1021/acs.nanolett.4c05326},
    issn = {15306992},
    pmid = {40037630}
}

@article{Feng2022EnhancedConcentration,
    title = {{Enhanced valley splitting in Si layers with oscillatory Ge concentration}},
    year = {2022},
    journal = {Physical Review B},
    author = {Feng, Yi and Joynt, Robert},
    number = {8},
    pages = {085304},
    month = {8},
    volume = {106},
    publisher = {American Physical Society},
    doi = {10.1103/PhysRevB.106.085304},
    issn = {24699969}
}

@article{Thorbeck2015FormationNanostructures,
    title = {{Formation of strain-induced quantum dots in gated semiconductor nanostructures}},
    year = {2015},
    journal = {AIP Advances},
    author = {Thorbeck, Ted and Zimmerman, Neil M},
    number = {8},
    month = {4},
    pages = {87107},
    volume = {5},
    doi = {10.1063/1.4928320},
    issn = {2158-3226}
}

@article{Windischmann1991IntrinsicFilms,
    title = {{Intrinsic stress in sputtered thin films}},
    year = {1991},
    journal = {Journal of Vacuum Science {\&} Technology A: Vacuum, Surfaces, and Films},
    author = {Windischmann, H.},
    number = {4},
    month = {7},
    pages = {2431--2436},
    volume = {9},
    doi = {10.1116/1.577295},
    issn = {0734-2101}
}

@article{Corley-Wiciak2023LatticeDevices,
  title = {Lattice Deformation at Submicron Scale: X-Ray Nanobeam Measurements of Elastic Strain in Electron Shuttling Devices},
  author = {Corley-Wiciak, C. and Zoellner, M.H. and Zaitsev, I. and Anand, K. and Zatterin, E. and Yamamoto, Y. and Corley-Wiciak, A.A. and Reichmann, F. and Langheinrich, W. and Schreiber, L.R. and Manganelli, C.L. and Virgilio, M. and Richter, C. and Capellini, G.},
  journal = {Phys. Rev. Appl.},
  volume = {20},
  issue = {2},
  pages = {024056},
  numpages = {12},
  year = {2023},
  month = {Aug},
  publisher = {American Physical Society},
  doi = {10.1103/PhysRevApplied.20.024056},
  url = {https://link.aps.org/doi/10.1103/PhysRevApplied.20.024056}
}

@inproceedings{ShimizuLocalEnhancement,
    title = {{Local mechanical-stress control (LMC): a new technique for CMOS-performance enhancement}},
    booktitle = {International Electron Devices Meeting. Technical Digest (Cat. No.01CH37224)},
    author = {Shimizu, A. and Hachimine, K. and Ohki, N. and Ohta, H. and Koguchi, M. and Nonaka, Y. and Sato, H. and Ootsuka, F.},
    pages = {1--19},
    publisher = {IEEE},
    isbn = {0-7803-7050-3},
    doi = {10.1109/IEDM.2001.979529}
}

@book{Ohring2001MaterialsEdition,
    title = {{Materials Science of Thin Films, 2nd Edition}},
    year = {2001},
    author = {Ohring, Milton.},
    edition = {2},
    pages = {},
    publisher = {Academic Press},
    isbn = {0125249756}
}

@article{McSkimin1953MeasurementWaves,
    title = {{Measurement of Elastic Constants at Low Temperatures by Means of High Frequency Ultrasonic Waves}},
    year = {1953},
    journal = {The Journal of the Acoustical Society of America},
    author = {McSkimin, H J},
    number = {4},
    month = {4},
    pages = {826},
    volume = {25},
    doi = {10.1121/1.1917687},
    issn = {0001-4966}
}

@article{Neyens2019MeasurementsArray,
    title = {{Measurements of Capacitive Coupling Within a Quadruple-Quantum-Dot Array}},
    year = {2019},
    journal = {Physical Review Applied},
    author = {Neyens, Samuel F and MacQuarrie, E R and Dodson, J P and Corrigan, J and Holman, Nathan and Thorgrimsson, Brandur and Palma, M and McJunkin, Thomas and Edge, L F and Friesen, Mark and Coppersmith, S N and Eriksson, M A},
    number = {6},
    month = {4},
    pages = {64049},
    volume = {12},
    doi = {10.1103/PhysRevApplied.12.064049},
    issn = {2331-7019}
}

@inproceedings{ItoMechanicalDesign,
    title = {{Mechanical stress effect of etch-stop nitride and its impact on deep submicron transistor design}},
    booktitle = {International Electron Devices Meeting 2000. Technical Digest. IEDM (Cat. No.00CH37138)},
    author = {Ito, S. and Namba, H. and Yamaguchi, K. and Hirata, T. and Ando, K. and Koyama, S. and Kuroki, S. and Ikezawa, N. and Suzuki, T. and Saitoh, T. and Horiuchi, T.},
    pages = {247--250},
    publisher = {IEEE},
    isbn = {0-7803-6438-4},
    doi = {10.1109/IEDM.2000.904303}
}

@book{Hibbeler2014MechanicsMaterials,
    title = {{Mechanics of materials}},
    year = {2014},
    author = {Hibbeler, R. C..},
    edition = {9},
    pages = {258--239},
    publisher = {Prentice Hall},
    isbn = {0133254429}
}

@article{Shi2006ModifyingImplantation,
    title = {{Modifying residual stress and stress gradient in LPCVD Si$_3$N$_4$ film with ion implantation}},
    year = {2006},
    journal = {Sensors and Actuators A: Physical},
    author = {Shi, Wendian and Zhang, Haixia and Zhang, Guobing and Li, Zhihong},
    number = {SPEC. ISS.},
    month = {8},
    pages = {352--357},
    volume = {130-131},
    publisher = {Elsevier},
    doi = {10.1016/J.SNA.2005.10.008},
    issn = {0924-4247}
}

@article{Corley-Wiciak2022NanoscaleDevice,
author = {Corley-Wiciak, Cedric and Richter, Carsten and Zoellner, Marvin H. and Zaitsev, Ignatii and Manganelli, Costanza L. and Zatterin, Edoardo and Schülli, Tobias U. and Corley-Wiciak, Agnieszka A. and Katzer, Jens and Reichmann, Felix and Klesse, Wolfgang M. and Hendrickx, Nico W. and Sammak, Amir and Veldhorst, Menno and Scappucci, Giordano and Virgilio, Michele and Capellini, Giovanni},
title = {Nanoscale Mapping of the {3D} Strain Tensor in a Germanium Quantum Well Hosting a Functional Spin Qubit Device},
journal = {ACS Applied Materials \& Interfaces},
volume = {15},
number = {2},
pages = {3119-3130},
year = {2023},
doi = {10.1021/acsami.2c17395},
URL = {https://doi.org/10.1021/acsami.2c17395}
}

@book{Levinshtein2001PropertiesSiGe,
  title={Properties of Advanced Semiconductor Materials: GaN, AIN, InN, BN, SiC, SiGe},
  author={Levinshtein, M.E. and Rumyantsev, S.L. and Shur, M.S.},
  isbn={9780471358275},
  lccn={00042262},
  series={A Wiley-Interscience publication},
  url={https://books.google.com/books?id=u26CpULkD_wC},
  year={2001},
  publisher={Wiley}
}

@book{Bradley2013PropertiesTemperatures,
    title = {{Properties of Selected Materials at Cryogenic Temperatures}},
    year = {2013},
    author = {Bradley, Peter and Radebaugh, Ray},
    publisher = {CRC Press, Boca Raton, FL},
    url = {https://tsapps.nist.gov/publication/get_pdf.cfm?pub_id=913059}
}

@article{Loss1998QuantumDots,
    title = {{Quantum computation with quantum dots}},
    year = {1998},
    journal = {Physical Review A},
    author = {Loss, Daniel and DiVincenzo, David P.},
    number = {1},
    month = {1},
    pages = {120},
    volume = {57},
    publisher = {American Physical Society},
    url = {https://journals.aps.org/pra/abstract/10.1103/PhysRevA.57.120},
    doi = {10.1103/PhysRevA.57.120},
    issn = {10941622},
    arxivId = {cond-mat/9701055}
}

@article{PaqueletWuetz2023ReducingWells,
    title = {{Reducing charge noise in quantum dots by using thin silicon quantum wells}},
    year = {2023},
    journal = {Nature Communications},
    author = {Paquelet Wuetz, Brian and Degli Esposti, Davide and Zwerver, Anne Marije J. and Amitonov, Sergey V. and Botifoll, Marc and Arbiol, Jordi and Vandersypen, Lieven M.K. and Russ, Maximilian and Scappucci, Giordano},
    number = {1},
    month = {12},
    pages = {1385},
    volume = {14},
    doi={10.1038/s41467-023-36951-w},
    publisher = {Nature Research},
    issn = {20411723},
    pmid = {36914637},
    arxivId = {2209.07242}
}

@misc{Frink2025ReducingStacking,
      title={Reducing strain fluctuations in quantum dot devices by gate-layer stacking}, 
      author={Collin C. D. Frink and Talise Oh and E. S. Joseph and Merritt P. Losert and E. R. MacQuarrie and Benjamin D. Woods and M. A. Eriksson and Mark Friesen},
      year={2025},
      eprint={2312.09235},
      archivePrefix={arXiv},
      url={https://arxiv.org/abs/2312.09235}, 
}

@article{Irene1976ResidualFilms,
    title = {{Residual stress in silicon nitride films}},
    year = {1976},
    journal = {Journal of Electronic Materials},
    author = {Irene, E. A.},
    number = {3},
    pages = {287--298},
    volume = {5},
    publisher = {Springer-Verlag},
    url = {https://link.springer.com/article/10.1007/BF02663273},
    doi = {10.1007/BF02663273/METRICS},
    issn = {1543186X}
}

@inproceedings{Guisbiers2006ResidualSubstrates,
    title = {{Residual Stresses in Ta, Mo, Al and Pd Thin Films Deposited by E-Beam Evaporation Process on Si and Si/SiO2 Substrates}},
    year = {2006},
    booktitle = {AIP Conference Proceedings},
    author = {Guisbiers, G and Overschelde, Olivier Van and Strehle, Steffen and Wautelet, M},
    pages = {317--324},
    publisher = {AIP},
    doi = {10.1063/1.2173565},
    issn = {0094243X}
}

@article{Ohya2014RoomResonators,
    title = {{Room temperature deposition of sputtered TiN films for superconducting coplanar waveguide resonators}},
    year = {2014},
    journal = {Superconductor Science and Technology},
    author = {Ohya, S and Chiaro, B and Megrant, A and Neill, C and Barends, R and Chen, Y and Kelly, J and Low, D and Mutus, J and O’Malley, P J J and Roushan, P and Sank, D and Vainsencher, A and Wenner, J and White, T C and Yin, Y and Schultz, B D and Palmstr{\o}m, C J and Mazin, B A and Cleland, A N and Martinis, John M},
    number = {1},
    month = {1},
    pages = {015009},
    volume = {27},
    url = {https://iopscience.iop.org/article/10.1088/0953-2048/27/1/015009},
    doi = {10.1088/0953-2048/27/1/015009},
    issn = {0953-2048}
}

@article{Burkard2023SemiconductorQubits,
    title = {{Semiconductor spin qubits}},
    year = {2023},
    journal = {Rev. Mod. Phys.},
    author = {Burkard, Guido and Ladd, Thaddeus D. and Pan, Andrew and Nichol, John M. and Petta, Jason R.},
    number = {2},
    month = {4},
    pages = {025003},
    volume = {95},
    publisher = {American Physical Society},
    doi = {10.1103/RevModPhys.95.025003},
    issn = {15390756},
    arxivId = {2112.08863}
}

@article{McJunkin:2022p7777,
      title={Si{G}e quantum wells with oscillating {Ge} concentrations for quantum dot qubits}, 
      author={Thomas McJunkin and Benjamin Harpt and Yi Feng and Merritt Losert and Rajib Rahman and J. P. Dodson and M. A. Wolfe and D. E. Savage and M. G. Lagally and S. N. Coppersmith and Mark Friesen and Robert Joynt and M. A. Eriksson},
  journal = {Nature Commun.},
  volume = {13},
  pages = {7777},
  numpages = {7},
  year = {2022},
  month = {Dec},
  doi = {10.1038/s41467-022-35510-z},
}

@article{Zwanenburg2013SiliconElectronics,
    title = {{Silicon quantum electronics}},
    year = {2013},
    journal = {Rev. Mod. Phys.},
    author = {Zwanenburg, Floris A and Dzurak, Andrew S and Morello, Andrea and Simmons, Michelle Y and Hollenberg, Lloyd C L and Klimeck, Gerhard and Rogge, Sven and Coppersmith, Susan N and Eriksson, Mark A},
    number = {3},
    month = {7},
    pages = {961--1019},
    volume = {85},
    publisher = {American Physical Society},
    url = {https://link.aps.org/doi/10.1103/RevModPhys.85.961},
    doi = {10.1103/RevModPhys.85.961}
}

@article{Hanson2007SpinsDots,
    title = {{Spins in few-electron quantum dots}},
    year = {2007},
    journal = {Rev. Mod. Phys.},
    author = {Hanson, R and Kouwenhoven, L P and Petta, J R and Tarucha, S and Vandersypen, L M K},
    number = {4},
    month = {10},
    pages = {1217--1265},
    volume = {79},
    publisher = {American Physical Society},
    url = {https://link.aps.org/doi/10.1103/RevModPhys.79.1217},
    doi = {10.1103/RevModPhys.79.1217}
}

@article{Losert2024StrategiesDisorder,
  title = {Practical strategies for enhancing the valley splitting in {Si/SiGe} quantum wells},
  author = {Losert, Merritt P. and Eriksson, M. A. and Joynt, Robert and Rahman, Rajib and Scappucci, Giordano and Coppersmith, Susan N. and Friesen, Mark},
  journal = {Phys. Rev. B},
  volume = {108},
  issue = {12},
  pages = {125405},
  numpages = {31},
  year = {2023},
  month = {Sep},
  publisher = {American Physical Society},
  doi = {10.1103/PhysRevB.108.125405},
  url = {https://link.aps.org/doi/10.1103/PhysRevB.108.125405}
}

@techreport{Thornton1989STRESS-RELATEDFILMS,
    title = {{STRESS-RELATED EFFECTS IN THIN FILMS}},
    year = {1989},
    booktitle = {Thin Solid Films},
    author = {Thornton, John A and Hoffman, D W},
    pages = {5--31},
    volume = {171}
}

@article{Capellini2014TensileProcess,
    title = {{Tensile Ge microstructures for lasing fabricated by means of a silicon complementary metal-oxide-semiconductor process}},
    year = {2014},
    journal = {Optics express},
    author = {Capellini, G. and Reich, C. and Guha, S. and Yamamoto, Y. and Lisker, M. and Virgilio, M. and Ghrib, A. and El Kurdi, M. and Boucaud, P. and Tillack, B. and Schroeder, T.},
    number = {1},
    month = {1},
    pages = {399},
    volume = {22},
    publisher = {Opt Express},
    url = {https://pubmed.ncbi.nlm.nih.gov/24515000/},
    doi = {10.1364/OE.22.000399},
    issn = {1094-4087},
    pmid = {24515000}
}

@article{Neyens2018TheWells,
    title = {{The critical role of substrate disorder in valley splitting in Si quantum wells}},
    year = {2018},
    journal = {Applied Physics Letters},
    author = {Neyens, Samuel F. and Foote, Ryan H. and Thorgrimsson, Brandur and Knapp, T. J. and McJunkin, Thomas and Vandersypen, L. M.K. and Amin, Payam and Thomas, Nicole K. and Clarke, James S. and Savage, D. E. and Lagally, M. G. and Friesen, Mark and Coppersmith, S. N. and Eriksson, M. A.},
    number = {24},
    month = {6},
    pages = {243107},
    volume = {112},
    publisher = {American Institute of Physics Inc.},
    doi = {10.1063/1.5033447},
    issn = {00036951},
    arxivId = {1804.01914}
}

@book{Touloukian1977ThermalSolids,
    title = {{Thermal expansion : nonmetallic solids}},
    year = {1977},
    author = {Touloukian, Y. S..},
    pages = {1786},
    publisher = {IFI/Plenum},
    isbn = {030667033X}
}

@article{Burke1962ThermalC,
  author  = {Burke, M.},
  title   = {Thermal expansion of ceramic materials at $-200^\circ$ to $0^\circ$C},
  journal = {Journal of the American Ceramic Society},
  year    = {1962},
  volume  = {45},
  pages   = {305--306}
}

@inproceedings{Corruccini1961ThermalLiterature,
    title = {{Thermal Expansion of Technical Solids at Low Temperatures: A Compilation From the Literature}},
    year = {1961},
    author = {Corruccini, Robert Joseph and Gniewek, John}
}

@article{Ylivaara2022ThermomechanicalDeposition,
    title = {{Thermomechanical properties of aluminum oxide thin films made by atomic layer deposition}},
    year = {2022},
    journal = {Journal of Vacuum Science {\&} Technology A},
    author = {Ylivaara, Oili M E and Langner, Andreas and Ek, Satu and Malm, Jari and Julin, Jaakko and Laitinen, Mikko and Ali, Saima and Sintonen, Sakari and Lipsanen, Harri and Sajavaara, Timo and Puurunen, Riikka L},
    number = {6},
    month = {4},
    pages = {62414},
    volume = {40},
    url = {https://doi.org/10.1116/6.0002095},
    doi = {10.1116/6.0002095},
    issn = {0734-2101}
}

@book{Freund2004ThinEvolution,
    title = {{Thin film materials : stress, defect formation, and surface evolution}},
    year = {2004},
    author = {Freund, L. B.. and Suresh, S..},
    pages = {86--151},
    publisher = {Cambridge University Press},
    isbn = {9780521822817}
}

@article{Marcks2025ValleyGermanium,
    title = {{Valley splitting correlations across a silicon quantum well containing germanium}},
    year = {2025},
    journal = {Nature Communications },
    author = {Marcks, Jonathan C. and Eagen, Emily and Brann, Emma C. and Losert, Merritt P. and Oh, Talise and Reily, J. and Wang, Christopher S. and Keith, Daniel and Mohiyaddin, Fahd A. and Luthi, Florian and Curry, Matthew J. and Zhang, Jiefei and Heremans, F. Joseph and Friesen, Mark and Eriksson, M. A.},
    number = {1},
    month = {12},
    pages = {11381},
    volume = {16},
    publisher = {Nature Research},
    doi = {10.1038/s41467-025-67325-z},
    issn = {20411723},
    pmid = {41353224}
}

@article{Lima2024ValleyDot,
    title = {{Valley splitting depending on the size and location of a silicon quantum dot}},
    year = {2024},
    journal = {Phys. Rev. Mater.},
    author = {Lima, Jonas R.F. and Burkard, Guido},
    number = {3},
    month = {3},
    pages = {036202},
    volume = {8},
    publisher = {American Physical Society},
    doi = {10.1103/PhysRevMaterials.8.036202},
    issn = {24759953},
    arxivId = {2310.17393}
}

@article{Hopcroft2010WhatSilicon,
    title = {{What is the Young's modulus of silicon?}},
    year = {2010},
    journal = {Journal of Microelectromechanical Systems},
    author = {Hopcroft, Matthew A. and Nix, William D. and Kenny, Thomas W.},
    number = {2},
    month = {4},
    pages = {229--238},
    volume = {19},
    doi = {10.1109/JMEMS.2009.2039697},
    issn = {10577157}
}

\end{document}